%% file: main.tex
\documentclass[preprint,12pt]{elsarticle}

\usepackage[utf8]{inputenc} 
\usepackage[T1]{fontenc}    
\usepackage{hyperref}       
\usepackage{url}            
\usepackage{booktabs}       
\usepackage{amsfonts}       
\usepackage{nicefrac}       
\usepackage{microtype}      
\usepackage{lipsum}
\usepackage{graphicx}       
\usepackage{algorithmic}
\usepackage{amsmath}
\usepackage{xcolor}
\usepackage{float}
\usepackage[linesnumbered,ruled,vlined]{algorithm2e}
\usepackage{multirow}
\usepackage{caption}
\usepackage{pgfplots}
\pgfplotsset{compat=1.18}

\graphicspath{{media/}}

\begin{document}

\title{Developing an open-source Framework for quantitative simulation of blood Flow and Tissue Motion for Ultrafast Doppler Ultrasound}

\author[inst1]{Qiang Fu}

\author[inst1]{Changhui Li\corref{cor1}}
\cortext[cor1]{Corresponding author}
\ead{chli@pku.edu.cn} 

\address[inst1]{College of Future Technology, Peking University, Beijing, China}

\begin{abstract}
Ultrafast power Doppler imaging (uPDI) has become a powerful tool for both research and clinical applications. However, existing simulation tools are insufficient for generating quantitatively accurate three-dimensional (3D) flow fields with tissue motion mimicking  in vivo conditions.   In this study, we present an open-source framework, named 3D-Fully Quantitative Flow (3D-FQFlow), to provide quan- titative modeling of 3D vascular hemodynamics with physiologically realistic  tissue motion for uPDI. The framework can perform quantitative modeling of both hemodynamics and tissue motion for either user-defined or clinical-derived vasculatures. Besides, it also integrates a GPU-accelerated image processing and reconstruction module. We demonstrate the performance of 3D-FQFlow using both synthetic vascular structures  and clinical datasets.  This framework could provide essential ground-truth simula- tion models to support the development, validation, and benchmarking of uPDI techniques.   The source code is freely available online at \url{https://github.com/FortuneOU/3D-FQFlow}.
\end{abstract}

\begin{keyword}
3D ultrasound simulation \sep Microvascular imaging \sep Tissue motion modeling \sep Open-source framework \sep Ultrafast power Doppler
\end{keyword}

\maketitle
\section{Introduction}
Since the 2010s, with the advent of ultrafast imaging technology and advances in clutter filter design, ultrafast Power Doppler Imaging (uPDI) has emerged as a powerful approach for microvascular imaging~\cite{2014Ultrafast,2017Noninvasive,2013Functional,2015Spatiotemporal}.Compared with conventional Doppler methods, uPDI transmits multiple angled plane waves with ultrafast frame rate and processes echo signals in parallel, providing markedly enhanced spatial resolution and  sensitivity to image low speed and small blood flows, broadening the clinical applications of power Doppler imaging~\cite{2011Ultrafast}.  

Nevertheless, constructing experimental phantoms capable of mimicking small-scale vascular structures with physiologically realistic hemodynamics and tissue motion remains strong challenging~\cite{2025In,0Improved},  particularly in 3D settings. Therefore, a robust simulation models is essential to exploration and validation of advanced uPDI techniques.  

Several tools have been developed to simulate blood flow and microbubble dynamics. Belgharbi et al.~\cite{Belgharbi2023_19} employed SIMUS~\cite{Garcia2022_20} to modele in vivo vascular networks as stacked 3D voxels and estimating velocity via a Poiseuille profile. Avdal et al.~\cite{Ekroll2023_21} proposed FLUST, which efficiently computes scatterer trajectories using streamlines but neglects tissue motion and realistic hemodynamics. Heiles and Chavignon et al.~\cite{Heiles2025_22} evaluated Ultrasound Localization Microscopy (ULM) algorithms using a Poiseuille-based model on the proprietary Verasonics platform. The BUFF framework~\cite{Lerendegui2022_23}, built upon Field II~\cite{Jensen1992_24,Jensen1999_25}, incorporates the Marmottant model and machine learning for vascular generation. Proteus~\cite{BlankenXX_26}, which employs k-Wave~\cite{Treeby2010_27}, better captures nonlinear propagation and microbubble scattering within vascular networks, but its computational demand---particularly for uPDI and 3D imaging hinders its implementation. Importantly, all existing platforms lack representations of perivascular tissue motion (cardiac and respiratory), which constitutes a major barrier to clinically relevant uPDI simulations. These limitations highlight the fundamental trade-off between simulation fidelity (hemodynamics, motion, and 3D realism) and computational tractability.  

Moreover, future machine learning in uPDI also demands large amount of anatomically and hemodynamically ground truth datasets for training. Therefore, there is an urgent demand for an integrated simulation platform capable of modeling vascular geometries, hemodynamic dynamics and physiological tissue motion~\cite{2022Super,2021Super,No0Functional,2024Pruning,Shin2024_18}.  

In this work, we propose the 3D-Fully Quantitative Flow (3D-FQFlow), a novel framework that uniquely integrates multiple physical modeling approaches, as well as optimized simulation and reconstruction algorithms, aiming to accurately simulate vascular hemodynamics and tissue motion with high fidelity. To better serve for uPDI society,  3D- FQFlow is released as an open-source MATLAB toolbox, available at \url{https://github.com/FortuneOU/3D-FQFlow}.

\section{3D-FQFlow framework}
\label{sec:framework}
The overall strategy of the proposed framework is illustrated in Figure 1, which synergistically integrates multiple available open-source hemodynamic simulation tools with optimized ultrasound simulation and imaging reconstruction algorithms.There are four modules, as follows.

\begin{figure}[H]
    \centering
    \includegraphics[width=1\linewidth]{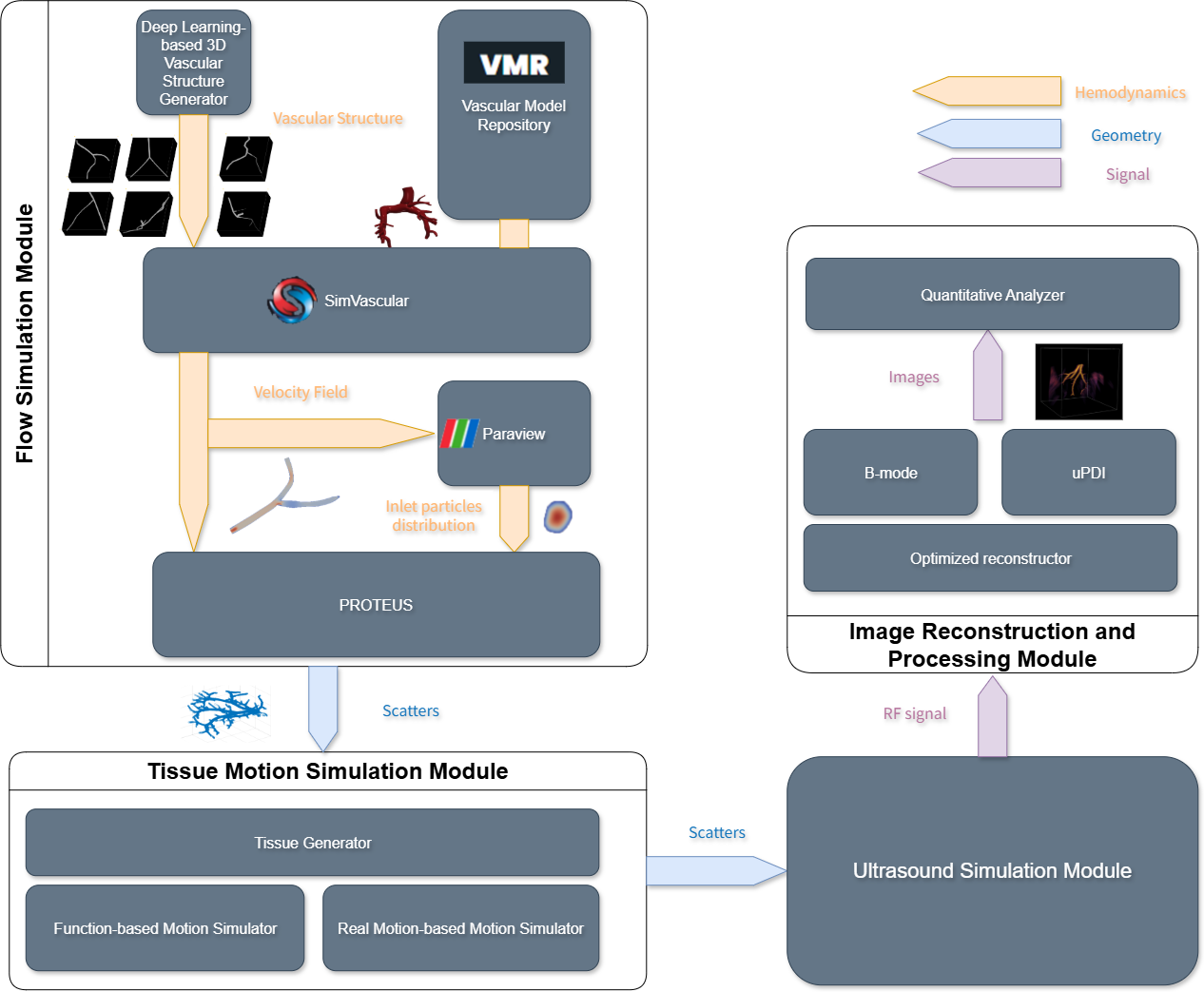}
    \caption{Architecture of the 3D-FQFlow simulator.}
    \caption*{The framework comprises four interconnected simulation modules to model blood vessels and perivascular tissues, followed by RF signal simulation and image reconstruction.}
    \label{fig:fig1}
\end{figure}

\begin{enumerate}
    \item \textbf{Flow Simulation Module:} Employs computational fluid dynamics (CFD) to simulate flow patterns in vascular structures from either medical imaging data or user-defined geometries~\cite{Jensen1999_25,GalarretaValverde2013_28,Adiv1985_30}.  
    \item \textbf{Tissue Motion Simulation Module:} Generates perivascular tissue models and enables 3D motion simulation through either parametric kinematic models or motion fields extracted from clinical data~\cite{Adiv1985_30}.  
    \item \textbf{ Ultrasound Simulation Module:} Implements a parallelized computing architecture to generate large-scale 3D radio frequency (RF) imaging data~\cite{Garcia2022_20}.  
    \item \textbf{Image Reconstruction and Processing Module:} Integrates an optimized 3D reconstruction algorithm, and multiple post-processing method for B-mode/uPDI images (2D/3D), and a quantitative analysis tool (MSE/PSNR/SSIM) to provide quantitative evaluation.  
\end{enumerate}

\subsection{Flow Simulation Module}
\label{sec:heading2}
As illustrated in Figure.\ref{fig:fig1}, the flow simulation module integrates a series of components to generate physiologically realistic blood flow data. The process begins with the creation of anatomically plausible vascular structures, proceeds with computational fluid dynamics (CFD) simulations to obtain hemodynamic fields, and subsequently derives the motion trajectories of intravascular scatterers. This pipeline collectively provides the essential ground truth hemodynamic data for the subsequent ultrasound simulation.
\subsubsection{Deep Learning-based 3D Vascular Structure Generator}
\label{sec:heading21}
The vascular structure generator utilizes a stochastic parametric Lindenmayer system (L-system) model~\cite{GalarretaValverde2013_28}, in which fractal-like vascular growth is simulated through iterative rewriting rules. The method is based on three assumptions: (i) bifurcation angles, vessel segment lengths, and diameters follow biomechanical constraints observed in vivo; (ii) vascular anomalies, such as aneurysms and stenoses, can be generated probabilistically; and (iii) vascular growth is spatially restricted by anatomical boundaries represented as 3D surfaces.  

The algorithm proceeds in three stages: (i) stochastic rewriting generates variable instruction strings; (ii) these instructions drive the construction of vascular skeletons via 3D transformations that simulate bifurcation processes; and (iii) convolution with Gaussian/step kernels produces CT/MRI-like intensity maps that mimic angiographic imaging.  

\begin{figure}[H]
    \centering
    \includegraphics[width=1\linewidth]{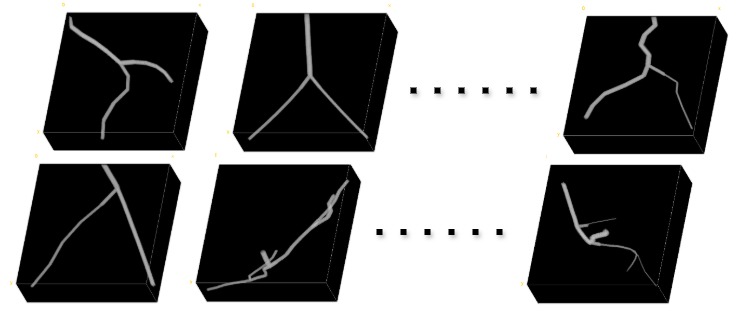}
    \caption{Representative vascular structures synthesized via the L-system model.}
    \caption*{Examples of 3D vascular structures generated by L-system, illustrating the ability of the L-system framework to synthesize versatile vascular architectures mimicking realistic conditions.}
    \label{fig:fig2}
\end{figure}

This approach ensures biomechanical realism—e.g., bifurcation angles within 35°–55° and diameters consistent with Murray’s law—while enabling user-defined 3D boundary constraints. By augmenting deterministic L-system models with stochastic operators, this method achieves statistically realistic vascular variability. Representative examples of the generated vascular structures are presented in Figure~\ref{fig:fig2}.

\subsubsection{SimVascular to Calculate Blood Flow}
\label{sec:heading22}
Once the vasculature is defined, the SimVascular~\cite{Updegrove2017_31,Lan2017_32},an open-source software platform for vascular modeling and flow calculation, is employed to segement vessels and and simulate the flow field. Its workflow includes segmentation  of vascular geometries from CT/MRI data ~\cite{Taylor1998_33,Marsden2015_34}, mesh generation~\cite{Sahni2008_35}, and hemodynamic simulations~\cite{Christian2001_36} using finite element or finite volume solvers. For the proposed 3D-FQFlow framework, the vascular skeleton generated by the L-system is first segmented, followed by mesh generation. Then, both inlets and outlets were identified, as well as setting inlet velocity distributions and outlet resistance values for every inlet and outlet, respectively. After that, computational fluid dynamics simulations were performed to obtain the intravascular flow field. Details of the implementation are provided in ~\ref{app:appA}.  

\subsubsection{ParaView to Assign Particle Injection}
\label{sec:heading23}
In our study, Doppler US echoes from blood is simulated using a large amount of particles as US scatterers~\cite{Garcia2022_20,Ekroll2023_21}. To extract particle injection statistics at vessel inlets, our framework uses ParaView~\cite{2005ParaView_39,Ayachit2015_40},an open-source platform built upon the Visualization Toolkit (VTK)—supports scalable data processing and rendering, to analyze velocity profiles across cross-sectional planes and determine particle distribution according to local inlet velocity, probability density functions derived from these velocity fields. Therefore, our framework can provide a physiologically meaningful basis for particle injection modeling (see ~\ref{app:appB}).  

\subsubsection{PROTEUS to Compute Particle Trajectory}
\label{sec:heading24}
To compute particle trajectories, the streamline of PROTEUS~\cite{BlankenXX_26} is employed. PROTEUS utilizes hemodynamic fields obtained from CFD simulations, and integrated trajectories of intravascular scatterers with MATLAB’s \texttt{ODE23} solver~\cite{1997The}. The resulting time-resolved particle trajectories are subsequently used as inputs for ultrasound scatterer modeling.  Details are further described in ~\ref{app:appC}.

\subsection{Tissue Motion Simulation Module}
\label{sec:heading3}
\begin{figure}[H]
    \centering
    \includegraphics[width=0.5\linewidth]{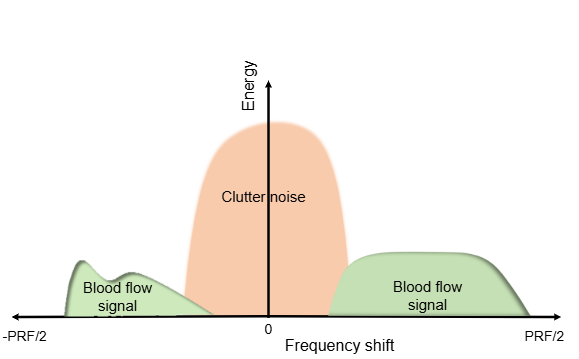}
    \caption{Schematic of Motion-Induced Frequency Shifts in Doppler Ultrasound}
    \caption*{The figure illustrates the frequency shift signals caused by motion during ultrasound Doppler imaging in real human tissues. The horizontal axis represents the frequency shift, while the vertical axis denotes the signal energy.}
    \label{fig:fig5}
\end{figure}

In practical clinical applications of uPDI, the frequency domain signals can generally be categorized into clutter noise(including signals from either stationary tissue or perivascular tissue motion) and blood flow signals, as illustrated in Figure.\ref{fig:fig5}. Among these, the signals generated by vascular tissue motion due to cardiac and respiratory activity are typically 20–30 dB stronger than the blood flow signals, which often results in the appearance of large, patchy artifacts (commonly referred to as "flash noise") in uPDI. This constitutes a primary challenge for Doppler ultrasound imaging. Furthermore, uPDI requires a longer ensemble size than conventional color Doppler imaging. Thus, accurate simulation of tissue motion is especially crucial for achieving clinically-representative uPDI simulations.

\subsubsection{Characterize Tissue Microstructure using Scatterers}
\label{sec:heading31}
To characterize tissue microstructure, this model employs randomly distributed point scatterers based on the framework of the  simulation of ultrasound (SIMUS) \cite{Garcia2022_20}, with acoustic field responses calculated through the principle of linear superposition. A density of approximately 83 scatterers per mm³ is utilized, and their reflection coefficients are assigned by sampling from a Rayleigh distribution with a mean value of 1. The detailed derivation of the formulas can be found in the ~\ref{app:appD}. In the example shown in Supplementary Video 1, approximately 4,000 points are located inside the vessel, accounting for about 0.16\%  of the total 2.5 million points.

\subsubsection{Modeling Tissue Motion for User-defined Steady Velocity Field}
\label{sec:heading32}

The tissue motion can be simulated for a user-defined steady velocity field $\vec{v}(x, y, z)$, in which the motion of scatterers are calculated by the following integral:
\begin{equation}
    \vec{P}(t) = \vec{P}_0 + \int_0^t \vec{v}(\vec{P}(s)) \, ds
\end{equation}
where
\begin{itemize}
    \item $\vec{P}(t) = (x(t), y(t), z(t))$ denotes the position vector of the particle at time $t$;
    \item $\vec{P}_0 = (x_0, y_0, z_0)$ represents the initial position of the particle;
    \item $\vec{v}(x, y, z) = (v_x(x, y, z), v_y(x, y, z), v_z(x, y, z))$ is the steady velocity field;
    \item $s$ is the dummy variable of integration, representing time from $0$ to $t$.
\end{itemize}
The key to this model is that all motion is governed by a defined velocity field. This allows for the separate computation of tissue and intravascular scatterer trajectories within the same framework. Moreover, by integrating the velocity field with respect to the known vascular geometry as a constraint, the motion of each intravascular scatterer can be calculated to ensure it remains confined within the vessel lumen throughout the simulation. The ultrasound simulation for each scatterer is also computed independently.  This approach ensures that all scatterers continuously satisfy the criteria described in\ref{app:appD}.

\subsubsection{Modeling Tissue Motion from Real Image Data}
\label{sec:heading33}
\begin{figure}[H]
    \centering
    \includegraphics[width=1\linewidth]{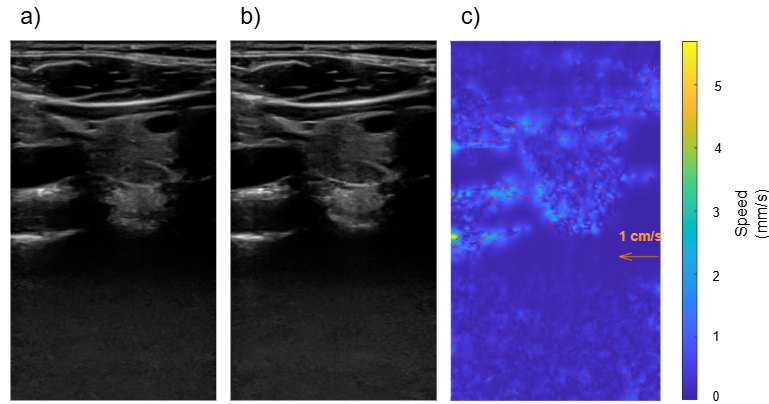}
    \caption{Tissue motion estimation of the clinical images Using optical flow}
    \caption*{a) and b) show two B-mode ultrasound images of the carotid artery region from a healthy volunteer, acquired using the clinical ultrasound system VINNO 5 (Vinno Technology (Suzhou) Co., Ltd.) with a time interval of one second. c) presents the tissue motion velocity vectors obtained by applying the optical flow method followed by median filtering to these two consecutive frames.}
    \label{fig:fig6}
\end{figure}

Tissue motion induced by physiological activities, such as respiration and heartbeat, generally exhibits periodic characteristics. The Tissue Motion Simulation Module is also capable of modeling tissue motion directly from real image data. To achieve this, the optical flow method~\cite{Adiv1985_30} is employed to extract motion patterns from clinical images, which are subsequently used to generate dynamic tissue phantoms. Figure~\ref{fig:fig6} illustrates the velocity fields obtained from real clinical ultrasound data.

\subsection{Ultrasound Simulation Module}
\label{sec:heading4}

One of the main challenges in performing 3D uPDI simulations lies in the  enormous computational cost for simulating hundreds of frames of complex  3D structures. This heavy computational burden is the primary reason why previous studies were either restricted to two-dimensional (2D) planes \cite{2021Super,2018Meshfree,2009Assessment,Ekroll2023_21,2016High,2018Color} or limited to only a small number of bubbles(scatterers) \cite{Heiles2025_22,Lerendegui2022_23,BlankenXX_26}. To address this limitation, 3D-FQFlow employs several optimizations based on the weak scattering assumption. These include considering only single scattering events while neglecting mutual interactions between scatterers, and leveraging distributed and parallel computing strategies—including GPU acceleration—at both the scatterer management and RF signal generation levels.

In conventional ultrasound simulation approaches, RF echoes from both blood flow and perivascular tissue scatterers are typically computed repeately for each frame, leading to substantial computational redundancy when dealing with the static tissue tissue whose components remain unchanged across frames. Therefore, for static tissue case, 3D-FQFlow employs an optimized strategy where scatterers representing blood flow and those located in the perivascular tissue are stored and managed separately. RF echoes from tissue scatterers outside the vessel are computed only for the initial frame. For subsequent frames, the total RF signal at each time point is efficiently obtained by simply adding the time-varying blood-flow RF echoes to the precomputed tissue background signals. This strategy enables much faster simulations under static tissue conditions, which constitutes one of the main advantages of 3D-FQFlow. In contrast, when tissue motion is present, tissue RF signals must be updated frame by frame. The detailed algorithmic workflow is provided in Algorithm~\ref{alg:rf-composition} in Appendix~\ref{app:appE}.

In this work, we employ an open-source tool, PFIELD \cite{Garcia2022_20}, a frequency-domain ultrasound simulation tool based on linear acoustic theory. PFIELD has been successfully integrated into SIMUS \cite{2016High,2018Meshfree,2018Color}. However, the PFIELD/SIMUS implementation generally requires substantial computational time, particularly in 3D scenarios\cite{GARCIA2024108169}, where generating a single frame could take hours.

To overcome this bottleneck, we propose an optimized PFIELD simulation method. Similar to the strategy in \cite{6841027}, the most computationally intensive operations—specifically, the matrix multiplications between propagation accumulation terms in the frequency domain and the incremental propagation terms from frequency stepping—are offloaded onto the GPU. Furthermore, based on the weak scattering assumption, scatterers are partitioned and processed via distributed computation. The partitioning strategy is dynamically adapted according to available GPU memory, which allows efficient execution even on machines equipped with relatively limited hardware (e.g., NVIDIA GeForce GTX 1050 with 2\,GB memory). The detailed procedure is described in Algorithm~\ref{alg:dist-acoustic-calc}. This approach enables high computational efficiency.

Table~\ref{tab:tab2} summarizes the computational performance of three simulation frameworks—Field~II, PFIELD, and optimized PFIELD—as the number of scatterers increases from $1 \times 10^{3}$ to $1 \times 10^{6}$. Simulations are conducted in an region of $4\,\mathrm{cm} \times 0.5\,\mathrm{cm} \times 4.5\,\mathrm{cm}$ using a $128$-element probe. Among the three methods, optimized PFIELD consistently achieves the best performance across all test scenarios,  which produces identical results to the standard PFIELD, requiring only 4,117 seconds for $1 \times 10^{6}$ scatterers. In contrast, the standard PFIELD implementation runs out of memory at the largest scale. These results clearly demonstrate that the optimized PFIELD substantially improves computational efficiency, particularly for large-scale 3D simulations.

\begin{table}[htbp]
  \centering
  \caption{Computation time(in second)for different simulators}
  \label{tab:tab2}
  \resizebox{\textwidth}{!}{
  \begin{tabular}{lcccc}
    \toprule
    simulators \textbackslash  number of scatterers  & $1\times10^3$ & $1\times10^4$ & $1\times10^5$ & $1\times10^6$ \\
    \midrule
    Field II  & 17  & 153  & 1517  & 15245  \\
    Existing PFIELD                        & 36  & 308  & 3412  & out of memory \\
    Optimized PFIELD               & 5   & 36   & 434   & 4117   \\
    \bottomrule
  \end{tabular}
  }
\end{table}

All simulations were performed on a workstation equipped with a 12-core Intel i7 CPU, an NVIDIA GeForce GTX 1050 GPU, and 64\,GB of RAM. The imaging setup employed an L11-4v probe with a seven-angle plane-wave transmission scheme. With these parameters, the optimized PFIELD required only about three hours to simulate RF signals for 200 frames.

\subsection{Image reconstruction and processing module}
\label{sec:headings5}

\subsubsection{Accelerated uPDI reconstruction}
\label{sec:headings51}

Under the assumption of the superposition principle of linear systems, Delay-and-Sum (DAS) processing for each spatial location can be performed independently without influencing other locations. Therefore, distributed computation is applied across all target spatial points, with grouping strategies dynamically determined according to the memory capacity of the runtime environment. In addition, the DAS matrices for each transmit angle are precomputed and stored persistently in memory, enabling newly acquired radio frequency (RF) data to be directly multiplied with the pre-stored matrices.  Using these strategies, we optimized the DAS computing within SIMUS~\cite{2020So,9593605}. This optimization removes the most time-consuming step, recomputing the delay-and-sum matrix for every frame, which is now calculated only once for multiple frames, thereby substantially reducing the reconstruction time.

\begin{figure}[H]
    \centering
    \includegraphics[width=1\linewidth]{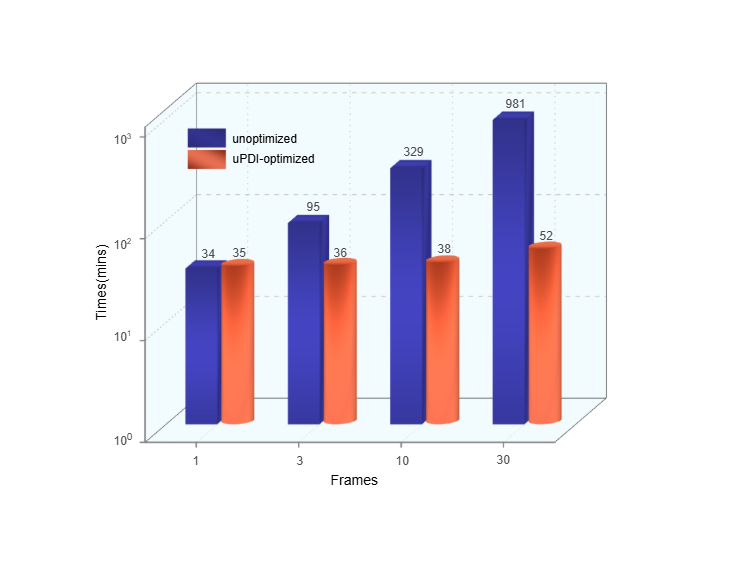}
    \caption{Comparison of computational time}
    \caption*{This figure presents a comparison of the computational time required by existing methods and the uPDI-optimized approach. The x-axis denotes the number of DAS frames, while the y-axis shows the elapsed time in minutes.}
    \label{fig:fig7}
\end{figure}

We conducted a comparative test of the computational time required by the conventional unoptimized reconstruction method and the optimized version for different numbers of frames, as illustrated in in Figure~\ref{fig:fig7}. The RF signals were simulated using a $32 \times 35$ matrix-array probe (Vermon, Tours, France), centered at 7.81\,MHz with an element pitch of 0.3\,mm. Five 2D tilted plane waves at 9\,MHz were transmitted. The imaging region was $2\,\mathrm{cm} \times 2\,\mathrm{cm} \times 2\,\mathrm{cm}$ with a spatial sampling interval of 0.2\,mm. For the conventional method, the computation time increases linearly with the number of frames: 34 minutes for one frame, 95 minutes for three frames, 329 minutes for ten frames, and 981 minutes for thirty frames. By contrast, the uPDI-optimized method required 35 minutes for one frame, 36 minutes for three frames, 38 minutes for ten frames, and only 52 minutes for thirty frames. These results clearly demonstrate that the optimized uPDI reconstruction substantially improves computational efficiency, with performance gains becoming more pronounced in multi-frame processing scenarios. All reconstruction experiments were executed on a PC equipped with a 20-core Intel i9 CPU, an NVIDIA GeForce GTX 1080Ti GPU, and 128\,GB of RAM.

\subsubsection{Post Processing}
\label{sec:headings52}
The 3D-FQFlow framework supports both brightness mode (B-mode) imaging and uPDI, with B‑mode images undergoing logarithmic compression and dynamic range display, and uPDI images processed with SVD filtering in addition to logarithmic compression and dynamic range display. Both imaging techniques are implemented in two-dimensional (2D) and three-dimensional (3D) formats, thereby enabling comprehensive visualization and analysis of anatomical structures and flow dynamics. In the subsequent sections (see Section~\ref{sec:headings62}), the results obtained using these imaging modalities are systematically presented and discussed in detail.

\subsubsection{Quantitative Analyzation}
\label{sec:headings53}
With the ground truth, the 3D-FQFlow framework further incorporates functions for quantitatively analyzing its reconstruction results. This evaluation employs three widely used objective metrics: mean squared error (MSE), peak signal-to-noise ratio (PSNR), and structural similarity index (SSIM). These metrics are applied to objectively assess the reconstruction quality and to provide a systematic evaluation of the algorithm's overall performance. The detailed results and analyses derived from this quantitative comparison are presented in Section~\ref{sec:headings61}.

\section{Results}
\label{sec:headings6}

Up to now, we have introduced a framework capable of quantitatively simulating both hemodynamics within vascular structures and tissue motion, as well as image reconstruction and processing.  It is important to note, however, that all modules in the framework can also work independently or be flexibly combined with other modules.  Several illustrative examples are provided in the following.

\subsection{Workflow Demonstration using Artificial 3D Vascular Structure}
\label{sec:headings64}
\begin{figure}[H]
    \centering
    \includegraphics[width=1\linewidth]{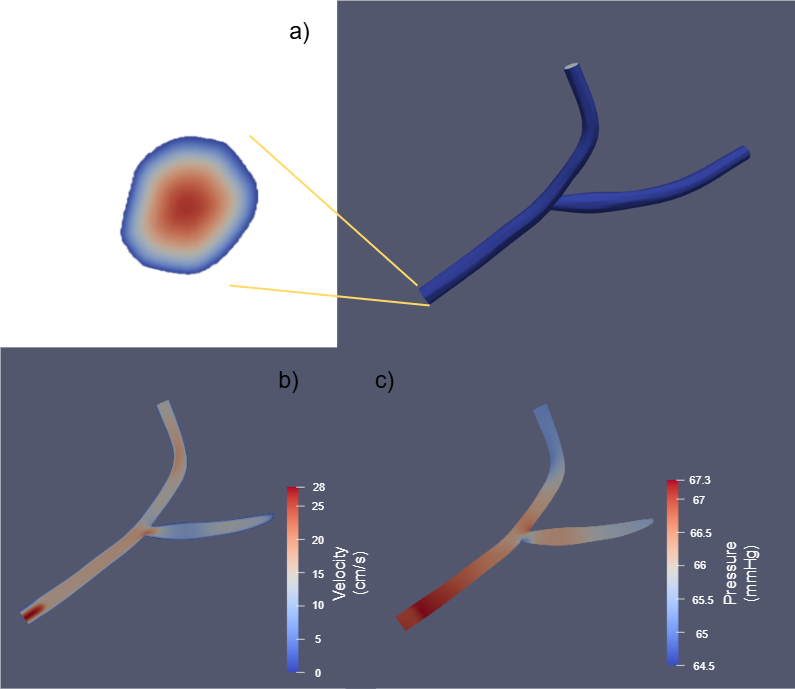}
    \caption{Structure, velocity field, and pressure field of vessel generated by L-system}
    \caption*{a) Overall structure and inlet projection. b) Velocity field on a slice. c) Pressure field on a slice.}
    \label{fig:fig4}
\end{figure}

Figure~\ref{fig:fig4} illustrates the results of a hemodynamic simulation within a vascular geometry generated using an L-system algorithm. Blood was modeled as a Newtonian fluid with a density of $1056\,\mathrm{kg}/\mathrm{m}^3$ and a kinematic viscosity of $3.27 \times 10^{-6}\,\mathrm{m}^2/\mathrm{s}$. The geometry consists of a single parent vessel bifurcating into two daughter branches. \\

To simulate flow velocity, the vascular  model was subsequently processed in \texttt{SimVascular} to extract centerlines and cross-sections for 3D model construction. An unstructured tetrahedral mesh was then created with \texttt{TetGen} module of SimVasculuar\cite{Updegrove2017_31}, and the Navier–Stokes equations were numerically solved using the finite-element-based solver to obtain the velocity field. Post-processing was performed in \texttt{ParaView}, including velocity field interpolation, planar slice extraction, and streamline visualization. In addition, particle tracking of 10,000 tracers was carried out using the \texttt{PROTEUS} framework. Adaptive step-size control and inflow resampling were employed to ensure numerical accuracy and statistical consistency, thereby enabling physiologically realistic reconstruction of 3D blood flow in branched vascular networks. The detailed computational procedures are described in Appendices~A,~B, and~C. \\

The simulated velocity field exhibits relatively high velocities at the inlet, while each bifurcated branch is characterized by higher central flow and reduced velocities near the vessel walls, consistent with canonical vessel flow patterns. The computed pressure field demonstrates a gradual decline along the vascular tree, with higher values maintained proximally near the inlet and progressive reductions at bifurcations.

\begin{figure}[H]
    \centering
    \includegraphics[width=1\linewidth]{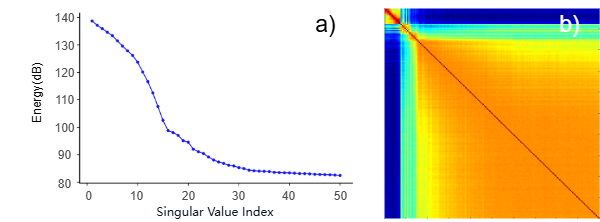}
    \caption{Schematic diagram of the components after SVD decomposition of vessel generated by L-system}
    \caption*{a) shows the line plot of the first 50 singular values after decomposition. b) respectively represent the correlation coefficient matrices of the left singular vectors (spatial modes) after taking the absolute values, following SVD decomposition.}
    \label{fig:fig13}
\end{figure}

Then, we go on to mimic a steady tissue motion by 3D-FQFlow based on the results presented above. Motion parameters are defined as $V_x$ = 3 mm/s, $V_y$ = 3 mm/s, $V_z$ = 3 mm/s. The corresponding singular value decomposition (SVD) results, which are derived in the image reconstruction module of our framework, are shown in the Figure.\ref{fig:fig13}.

\begin{figure}[H]
    \centering
    \includegraphics[width=1\linewidth]{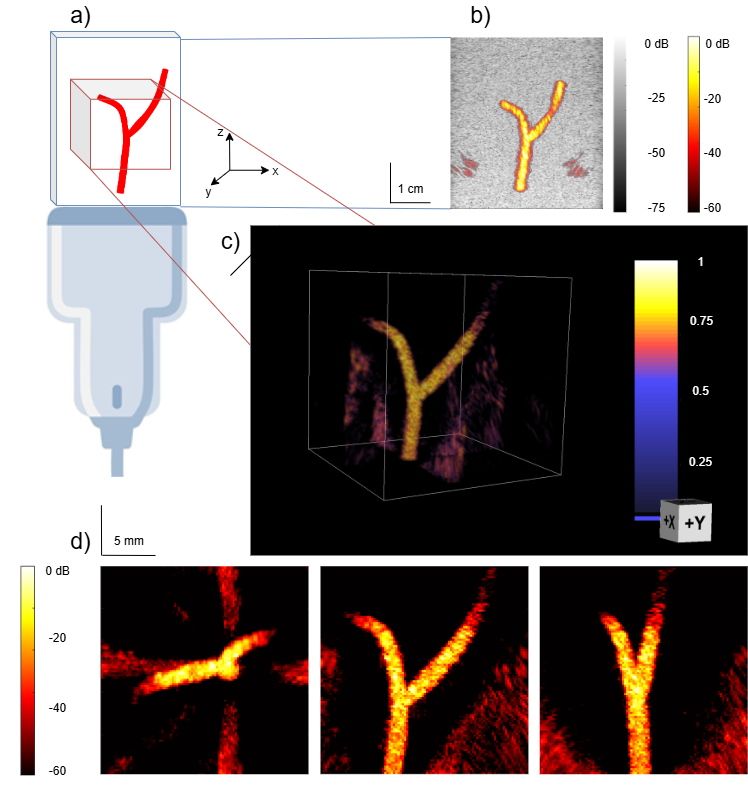}
    \caption{B-mode and uPDI imaging of vascular structure generated by deep learning-based 3D Vascular Structure Generator}
    \caption*{
        a) Schematic diagram of the imaging procedure. 
        b) Fused B-mode and uPDI image acquired by the L11-4v linear array probe. 
        c) 3D uPDI imaging result rendered with Volview software. 
        d) Maximum intensity projections of the 3D uPDI data along different planes.
    }
    \label{fig:fig9}
\end{figure}

Figure~\ref{fig:fig9} presents the B-mode and uPDI imaging results of the vascular structure generated in Figure~\ref{fig:fig4}, obtained using both the L11-4v linear probe and the Vermon matrix-array probe, respectively. Figure~\ref{fig:fig9}(a) shows a schematic diagram of the imaging alignment. Figure~\ref{fig:fig9}(b) displays the fused image by superposing uPDI image onto B-mode result acquired with the L11-4v linear array, covering an imaging area of $4\,\mathrm{cm} \times 4.5\,\mathrm{cm}$, with dynamic ranges of 75\,dB for B-mode and 60\,dB for uPDI, respectively. Figure~\ref{fig:fig9}(c) shows the 3D uPDI result visualized using \texttt{VolView}, covering an imaging volume of $2\,\mathrm{cm} \times 2\,\mathrm{cm} \times 2\,\mathrm{cm}$. Figure~\ref{fig:fig9}(d) presents the maximum-intensity projections (MIPs) of the 3D uPDI data onto the $x\text{-}y$, $x\text{-}z$, and $y\text{-}z$ planes, with a dynamic range of 60\,dB. Supplementary Video~1 provides an additional dynamic visualization of these results.

The imaging results in Figure~\ref{fig:fig9} confirm the validity of the proposed framework: both the linear array and the matrix-array probes produced uPDI images that closely match the vascular structures. In addition, quantitative evaluation yielded a mean squared error (MSE) of 0.0027, a peak signal-to-noise ratio (PSNR) of 25.61, and a structural similarity index (SSIM) of 0.902, further supporting the fidelity of the reconstructed flow patterns. More dynamic visualization of the 3D imaging features in Figure~\ref{fig:fig9} are provided in Supplementary Video~1.\\

It is worthy to note that using  \texttt{SimVascular} for flow simulations of more complex vascular geometries, intensive and iterative technical adjustments are required with long learning curve. These include skeleton extraction, mesh generation, and boundary-condition specification.  Nevertheless, this challenge highlights an important direction for future work - namely, the development of automated or semi-automated pipelines that can streamline these steps and reduce user intervention.

\subsection{Simulation of Tissue Motions}
\label{sec:headings61}
\begin{figure}[H]
    \centering
    \includegraphics[width=1\linewidth]{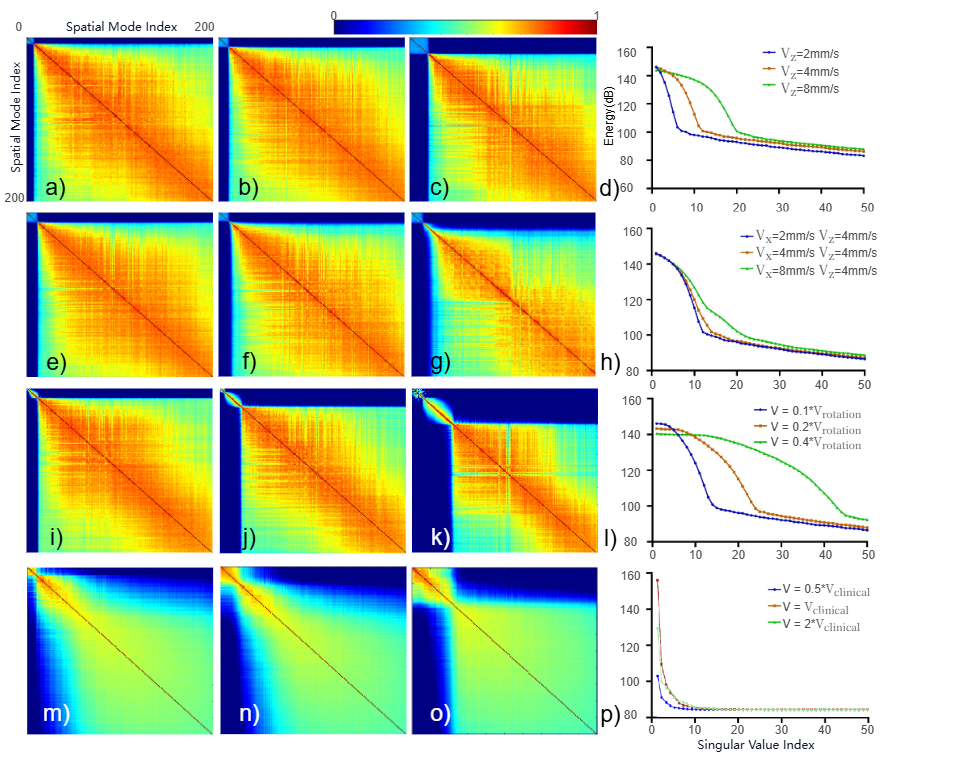}
    \caption{Schematic diagram of the components after SVD decomposition under different tissue motions}
    \caption*{a), b), and c) respectively represent the correlation coefficient matrices of the left singular vectors (spatial modes) after taking the absolute values, following SVD decomposition, when the perivascular tissue moves at $V_z$ = 2 mm/s, $V_z$ = 4 mm/s, and $V_z$ = 8 mm/s. d) shows the line plot of the first 50 singular values after decomposition to a), b), and c); 
        e), f), and g) show the correlation coefficient matrices of the absolute values of the left singular vectors (spatial modes) after SVD decomposition, corresponding to the cases where the perivascular tissue moves at $V_x$ = 2 mm/s, $V_x$ = 4 mm/s, and $V_x$ = 8 mm/s, respectively, with a fixed z-direction velocity of 4 mm/s. h) presents the line plot of the first 50 singular values obtained after the decomposition to e), f), and g); 
        i), j), and k) show the correlation coefficient matrices of the absolute values of the left singular vectors (spatial modes) after SVD decomposition, corresponding to rotational motions of the perivascular tissue, with the velocity field defined as \(\mathbf{V} = (0,\,-\omega,\,0)\times[(x,\,y,\,z)-(0,\,0,\,2.5)]\)cm/s, \(\omega = 1\,\text{rad/s}\), and is scaled by factors of 0.1, 0.2, and 0.4, respectively. l) presents the line plots of the first 50 singular values after decomposition corresponding to i, j, and k;
        m), n), and o) show the correlation coefficient matrices of the absolute values of the left singular vectors (spatial modes) after SVD decomposition, corresponding to cases where the actual tissue motion velocity is scaled by factors of 0.5, 1, and 2, respectively. p) presents the line plots of the first 50 singular values after decomposition corresponding to m), n), and o).}
    \label{fig:fig11}
\end{figure}

The proposed framework includes a unique tissue-motion simulation module, which can also be implemented independently. As an illustrative example, Figure~\ref{fig:fig11} shows the results of 2D simulation of mimicing tissue motion performed on a rabbit kidney vasculature. The simulation modeled four distinct perivascular tissue motion patterns: (1) translation along the $z$-axis, (2) translation along a diagonal path in the $x$-$z$ plane, (3) rotation about a central point, and (4) complex motion derived from clinical tissue-motion data. Influence of tissue motion on the spatial mode vectors and singular values obtained from SVD were analyzed at three different motion velocities. The results indicate that for motion along the $z$-axis, rotational motion about the central point, and motion derived from clinical data, the magnitude of the frequency shift induced by tissue motion increases significantly with rising velocity. In contrast, diagonal motion in the $x$-axis direction introduces measurable alterations in the SVD results, but the overall impact remains comparatively limited.

\begin{figure}[H]
    \centering
    \includegraphics[width=1\linewidth]{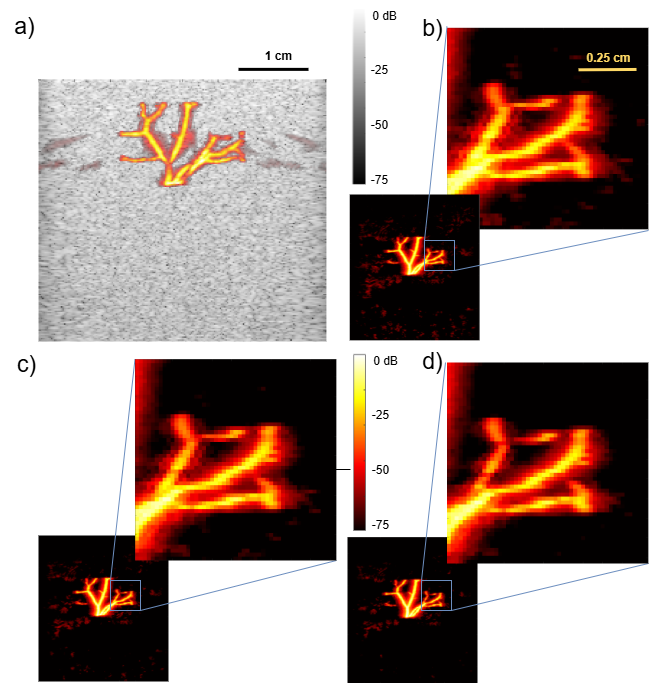}
    \caption{uPDI results for parivascular tissue motion under different velocities}
    \caption*{a) B-mode image of the initial frame;
b) uPDI result when the actual tissue motion velocity is scaled by a factor of 0.5;
c) uPDI result when the actual tissue motion velocity is scaled by a factor of 1;
d) uPDI result when the actual tissue motion velocity is scaled by a factor of 2.}
    \label{fig:fig12}
\end{figure}

Figure~\ref{fig:fig12} shows the simulated imaging results obtained by scaling the clinical tissue-motion velocity by factors of 0.1, 0.2, and 0.4. The images illustrate the effect when the cutoff threshold of the clutter filter is positioned at the location indicated by the arrow in Figure~\ref{fig:fig11}(p), with a system dynamic range of 75\,dB. The results demonstrate that as the tissue-motion velocity increases, the continuity of small-vessel depiction in the reconstructed images progressively deteriorates.

\begin{table}[htbp]
\centering
\caption{MSE, PSNR, SSIM for different objects}
\label{tab:tab3}
  \resizebox{\textwidth}{!}{
\begin{tabular}{lccc}
\toprule
Object & MSE & PSNR & SSIM \\
\midrule
$V_z = 2\,\mathrm{mm/s}$ & 0.0023424 & 26.3033 & 0.95413 \\
$V_z = 4\,\mathrm{mm/s}$ & 0.0031058 & 25.0782 & 0.94638 \\
$V_z = 8\,\mathrm{mm/s}$ & 0.0048677 & 23.1268 & 0.93302 \\
$V_z = 4\,\mathrm{mm/s}, V_x = 2\,\mathrm{mm/s}$ & 0.0034401 & 24.6343 & 0.94276 \\
$V_z = 4\,\mathrm{mm/s}, V_x = 4\,\mathrm{mm/s}$ & 0.0047736 & 23.2115 & 0.9324 \\
$V_z = 4\,\mathrm{mm/s}, V_x = 8\,\mathrm{mm/s}$ & 0.0074532 & 21.2766 & 0.92216 \\
$V = 0.1 \times V_{\text{rotation}}$ & 0.0016793 & 27.7487 & 0.97081 \\
$V = 0.2 \times V_{\text{rotation}}$ & 0.0023528 & 26.2842 & 0.96465 \\
$V = 0.4 \times V_{\text{rotation}}$ & 0.0029322 & 25.3281 & 0.95912 \\
$V = 0.5 \times V_{\text{clinical}}$ & 0.0036448 & 25.1231 & 0.96332 \\
$V = V_{\text{clinical}}$ & 0.0041845 & 24.3960 & 0.95644 \\
$V = 2 \times V_{\text{clinical}}$ & 0.0057491 & 23.2729 & 0.95198 \\
\bottomrule
\end{tabular}
}
\end{table}

Table~\ref{tab:tab3} summarizes the quantitative evaluation results of the tissue-motion analysis on the maximum intensity  projection (MIP) images, including mean squared error (MSE), peak signal-to-noise ratio (PSNR), and structural similarity index (SSIM). 

For unidirectional motion along the $z$-axis ($V_z = 2/4/8\,\mathrm{mm}/\mathrm{s}$), MSE increases with velocity, while PSNR and SSIM decrease accordingly. When $V_z$ is fixed at $4\,\mathrm{mm}/\mathrm{s}$ and the transverse velocity component varies ($V_x = 2/4/8\,\mathrm{mm}/\mathrm{s}$), elevated transverse velocity exacerbates MSE growth and accelerates degradation of PSNR and SSIM. 

Scaling experiments for rotational motion ($V_{\text{rotation}}$) and clinical motion ($V_{\text{clinical}}$) confirm that low-speed motion (0.5$\times$) produces substantially higher fidelity than high-speed motion (2$\times$). Notably, rotation motion achieves the highest SSIM value (97081) at 0.1$\times$ scaling. Furthermore, Supplementary Video~2 provides a dynamic visualization of B-mode and uPDI images for a 3D phantom undergoing motion at $4\,\mathrm{mm}/\mathrm{s}$ along the positive $z$-axis.

\subsection{Image Reconstruction of Real Vascular Structures}
\label{sec:headings62}

This section demonstrates the applicability and robustness of the proposed imaging framework when applied to realistic complex anatomical geometries, including an animal organ vasculature (rabbit kidney) and a human clinical data from a post-Glenn patient of open source data\cite{Troianowski2011_38}. Therefore, we can assess the image reconstruction performance of the proposed framework under physiologically and clinically relevant vascular morphologies. \\

\begin{figure}[H]
    \centering
    \includegraphics[width=1\linewidth]{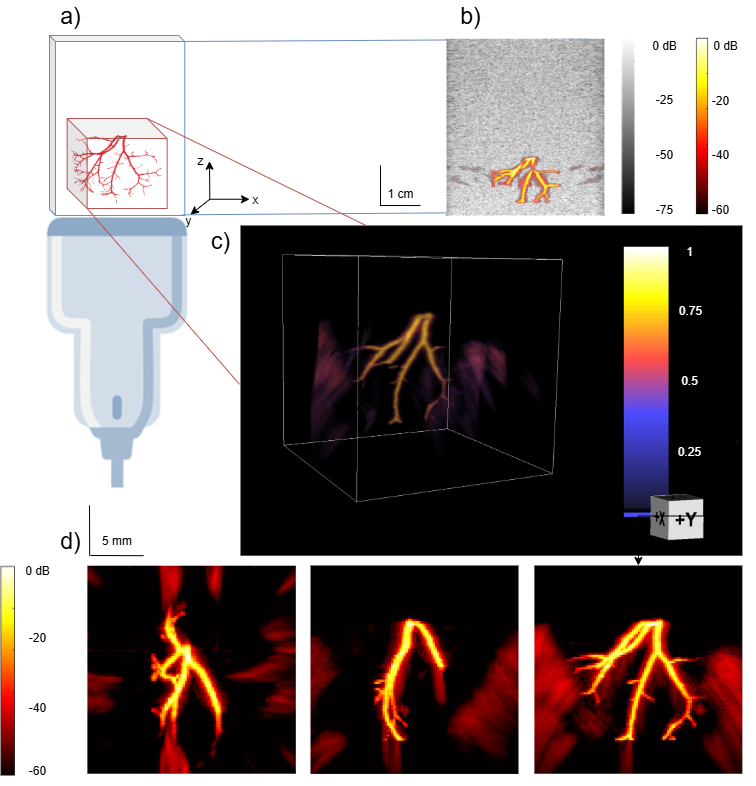}
    \caption{B-mode and uPDI imaging of a rabbit kidney using linear and matrix-array probes}
    \caption*{a) Schematic diagram of the imaging procedure. 
        b) Fused B-mode and uPDI image acquired by the linear array probe. 
        c) 3D uPDI imaging result rendered with Volview software. 
        d) Maximum intensity projections of the 3D uPDI data along different planes.}
    \label{fig:fig8}
\end{figure}

Figure~\ref{fig:fig8} presents the B-mode and uPDI imaging results of a rabbit kidney obtained using both the L11-4v linear probe and the Vermon matrix-array probe. Figure~\ref{fig:fig8}(a) shows a schematic diagram of the imaging setup. Figure~\ref{fig:fig8}(b) displays the fused B-mode and uPDI image acquired with thelinear array, covering an imaging area of $4\,\mathrm{cm} \times 4.5\,\mathrm{cm}$, with a dynamic range of 75\,dB for B-mode and 60\,dB for uPDI; the reconstruction time for the data was approximately 25 minutes. Figure~\ref{fig:fig8}(c) shows the 3D uPDI imaging result visualized using the \texttt{VolView} software \cite{2023Interactive}, covering an imaging volume of $2\,\mathrm{cm} \times 2\,\mathrm{cm} \times 2\,\mathrm{cm}$, with a reconstruction time of about 125 minutes for the matrix-array data. Figure~\ref{fig:fig8}(d) presents the maximum-intensity projections (MIPs) of the 3D uPDI data on the $x\text{-}y$, $x\text{-}z$, and $y\text{-}z$ planes, with a dynamic range of 60\,dB. Supplementary Video~3 provides a dynamic visualization of these results.

\begin{figure}[H]
    \centering
    \includegraphics[width=1\linewidth]{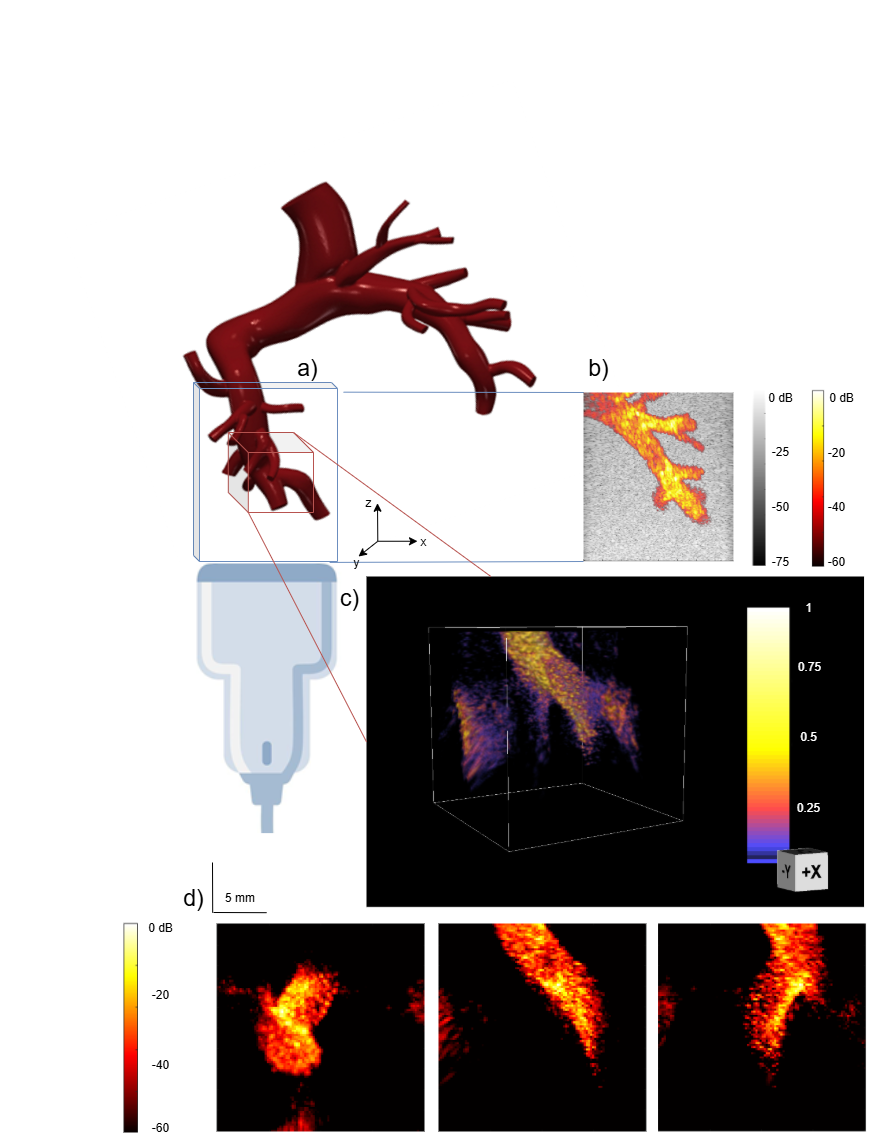}
    \caption{B-mode and uPDI imaging of the clinical branch pulmonary arteries (left half) of a post-Glenn patient}
    \caption*{
        a) Schematic diagram of the imaging procedure. 
        b) Fused B-mode and uPDI image acquired by the L11-4v linear array probe. 
        c) 3D uPDI imaging result rendered with Volview software. 
        d) Maximum intensity projections of the 3D uPDI data along different planes.
    }
    \label{fig:fig10}
\end{figure}

Figure~\ref{fig:fig10} presents the B-mode and uPDI imaging results of the clinical branch pulmonary arteries (left half) of a post-Glenn patient, corresponding to the vascular structure shown in Figure~\ref{fig:fig3}(b). The results were obtained using both the L11-4v linear probe and the Vermon matrix-array probe, respectively. Figure~\ref{fig:fig10}(a) shows a schematic diagram of the imaging setup. Figure~\ref{fig:fig10}(b) displays the fused B-mode and uPDI image acquired with the 2D L11-4v linear array, covering an imaging area of $4\,\mathrm{cm} \times 4.5\,\mathrm{cm}$, with dynamic ranges of 75\,dB for B-mode and 60\,dB for uPDI, respectively. Figure~\ref{fig:fig10}(c) shows the 3D uPDI result visualized using the \texttt{VolView} software, covering an imaging volume of $2\,\mathrm{cm} \times 2\,\mathrm{cm} \times 2\,\mathrm{cm}$. Figure~\ref{fig:fig10}(d) presents the maximum-intensity projections (MIPs) of the 3D uPDI data onto the $x\text{-}y$, $x\text{-}z$, and $y\text{-}z$ planes, with a dynamic range of 60\,dB. \\

Furthermore, Supplementary Video~3 and Supplementary Video~4 provide more dynamic visualization of the 3D imaging features presented in Figures~\ref{fig:fig8} and~\ref{fig:fig10}.

\begin{table}[htbp]
\centering
\caption{MSE, PSNR, SSIM for different biological structures}
\label{tab:tab4}
  \resizebox{\textwidth}{!}{
\begin{tabular}{lccc}
\toprule
Object & MSE & PSNR & SSIM \\
\midrule
rabbit kidney & 0.0011462 & 29.4075 & 0.95104 \\
clinical branch pulmonary arteries & 0.004325 & 23.6395 & 0.85001 \\
\bottomrule
\end{tabular}
}
\end{table}

The results in Table~\ref{tab:tab4} provide quantitative assessments of 3D vascular imaging features, further validating the robustness of the proposed framework in complex vascular morphologies.

\section{Discussion}
\label{sec:headings7}
The 3D-FQFlow framework was developed as an open-source platform that integrates 3D hemodynamic modeling with tissue-motion simulation. In addition, the framework substantially reduces the computational time for large-scale 3D uPDI simulation (e.g., approximately 3 hours for 200 frames with one million scatterers). This efficiency enables researchers to systematically quantify the impact of motion artifacts on uPDI under conditions with known ground truth (i.e., vessel structure, velocity field, and tissue displacement). \\

The current version, however, has several limitations, primarily related to hemodynamic simplifications. Although non-uniform flow is supported, the constant-velocity assumption and rigid-vessel model neglect important physiological complexities such as pulsatile flow and vessel compliance. In addition, the scatterer model accounts only for single-component linear properties (e.g., echo coefficients) while excluding nonlinear oscillation effects of microbubbles. \\

Moreover, when employing \texttt{SimVascular} for vascular simulations of complex geometries, the workflow still requires technically intensive and iterative steps, including skeleton extraction, mesh refinement, and boundary-condition specification. These operations impose a non-negligible learning curve and reduce the overall accessibility of the current pipeline. Nevertheless, this limitation highlights an important future direction: the development of automated or semi-automated modules to streamline these operations, thereby lowering the entry barrier while enhancing reproducibility. \\

\section{Conclusion}
\label{sec:headings8}

In the following Table.\ref{tab:tab1}, we compared our framework with current simulation tools used by uPDI. Our framework shows unique advantage.\\
\begin{table}[htbp]
\centering
\caption{Comparison of ultrasound simulation methods.}
\label{tab:tab1}
\resizebox{\textwidth}{!}{
\begin{tabular}{lcccccc}
\toprule
 & \textbf{SIMUS} & \textbf{FLUST} & \textbf{Verasonics} & \textbf{BUFF} & \textbf{Proteus} & \textbf{3D-FQFlow} \\
\midrule
Ultrasound simulation methods & PFIELD & COLE & Verasonics & Field II & k-wave & optimized PFIELD \\
Dimensionality  & 2D,3D & 2D & 2D & 2D,3D & 2D,3D  & 2D,3D  \\
Structure generation using ML & no & no & no & yes & no & yes \\
Realistic structures & yes & no & no & no & yes & yes \\
Simulation of tissue motion & no & * & no & no & no & yes \\
Computational speed & * & fast & * & * & slow & fast \\
Nonlinearity & no & no & no & partial & yes & no \\
Variable blood flow velocity & no & no & no & no & no & yes \\
Non-Poiseuille flow simplification & no & no & no & yes & yes & yes \\
\end{tabular}
}
\end{table}

3D-FQFlow establishes a framework for ultrasound power Doppler imaging (uPDI) that supports tissue motion simulation alongside 3D non-uniform blood flow simulation. Its open-source architecture (available at \url{https://github.com/FortuneOU/3D-FQFlow}) and distributed computing optimizations (including the GPU-accelerated PFIELD simulator and memory-partitioned reconstruction) overcome the computational bottlenecks inherent in traditional tools. This breakthrough enables the feasible generation of hundreds of frames of 3D ultrasound data with corresponding ground truth (vessel structure, velocity field, tissue displacement). The framework has demonstrated its value in motion artifact analysis, vascular structure imaging, and algorithm validation.

Future development will focus on:

\begin{enumerate}
    \item Flow simulation module to handle more complex vascular structures.
    \item Enhancing hemodynamic fidelity (e.g., incorporating pulsatile flow and deformable vessel models).
    \item Extending the scatterer model (to include distributions of scatterer size and acoustic properties).
    \item Deepening machine learning synergy (leveraging simulated data to train motion-artifact-resistant models).

\end{enumerate}

Although improvements in the fidelity and convenience of hemodynamic modeling are still required and simulations of nonlinear acoustic effects have not yet been incorporated, these developments are expected to be integrated in subsequent versions. Overall, 3D-FQFlow is a preliminary but useful step toward standardized simulation framework for uPDI research. We expect it to complement ongoing advances in imaging parameter optimization, clutter filter design, and learning-based ultrasound image analysis.

\section*{Acknowledgment}
This research was supported by the following grants: the National Key R\&D Program of China (No.2023YFC2411700); the Beijing Natural Science Foundation (No.7232177). 
The authors acknowledge the developers and contributors of multiple open-source software packages, including the Matlab Ultrasound Toolbox (MUST), PROTEUS, SimVascular, and ParaView, which were essential for ultrasound simulation, flow tracking, hemodynamic analysis, and visualization in this study.
The authors also acknowledge the use of data from the Vascular Model Repository (www.vascularmodel.com ), supported by the National Library of Medicine and the National Heart, Lung, and Blood Institute, in accordance with its licensing terms.

\appendix

\input{appendix_content.tex}

\bibliographystyle{elsarticle-num}
\bibliography{references}

\end{document}

%% file: appendix_content.tex
\section{SimVascular workflow} \label{app:appA}
\begin{figure}[H]
    \centering
    \includegraphics[width=1\linewidth]{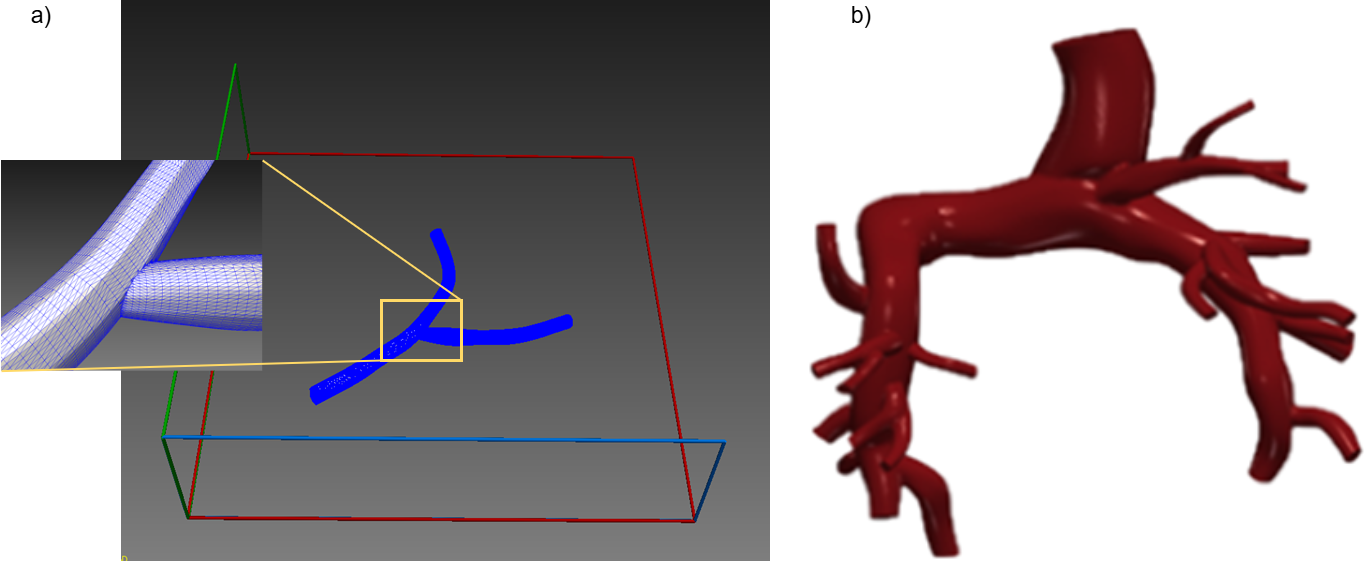}
    \caption{Simvascular}
    \caption*{a) Unstructured tetrahedral meshes. b) Multi-branched pulmonary arteries and superior vena cava structure of a Glenn patient.}
    \label{fig:fig3}
\end{figure}

In this study, the graphical user interface of SimVascular.Window.05.2023 was employed to extract the centerlines and cross-sections from the Deep Learning-based 3D Vascular Structure Generator, enabling the construction of three-dimensional vascular models.The workflow was as follows:\\
\begin{enumerate}
    \item {Data importation:} Import the 3D vascular geometry generated by the Deep Learning model into the SimVascular project workspace.
    \item {Centerline extraction:} Extract vessel centerlines using the automated path tool. Manual correction was applied at bifurcations.
    \item {Cross-sectional generation:} Generate orthogonal cross-sections along each centerline to delineate lumen contours.
    \item {Solid modeling:} Interpolate between cross-sections and apply smoothing to construct vascular solid surfaces.
    \item {Mesh generation:} Generate unstructured tetrahedral meshes via TetGen\cite{Si2015_37} (see Figure.~\ref{fig:fig3}.a). Mesh density was iteratively adjusted to balance precision and efficiency.
    \item {Simulation:} Navier–Stokes equations were solved with \texttt{svSolver} (FEM-based), using case-specific inlet/outlet boundary conditions.
\end{enumerate}

 In addition, the official SimVascular website (\url{https://www.vascularmodel.com}) offers a variety of pre-constructed vascular flow models in different morphologies for user access; an example\cite{Troianowski2011_38} of multi-branched pulmonary arteries and superior vena cava structure of a Glenn patient is illustrated Figure.\ref{fig:fig3}b.\\

\section{Paraview workflow} \label{app:appB}
 The platform accommodates diverse data structures including structured grids, unstructured grids, and volumetric data, while implementing a comprehensive workflow through its modular pipeline architecture. This workflow encompasses data importation, preprocessing (filtering/clipping/sampling), visualization mapping (volume rendering/surface reconstruction/isosurface extraction), and result output. Its MPI-based parallel processing architecture significantly enhances distributed computation and real-time rendering efficiency for large-scale scientific datasets.\\
In this investigation, post-processing and visualization were performed in ParaView 5.13 through the following steps:
\begin{enumerate}
    \item Load simulation results in VTK format.
    \item Apply ``Cell Data From Point Data'' to interpolate nodal velocities onto cells.
    \item Use the ``Slice'' filter to extract planar sections at anatomical locations.
    \item Use the ``Stream Tracer'' with inlet seeds to visualize particle trajectories.
    \item Render and export high-resolution streamline and velocity plots in TIFF/PDF format.
\end{enumerate}

\section{PROTEUS workflow} \label{app:appC}
The algorithm incorporates two fundamental assumptions: (1) During the temporal scale encompassing ultrasonic pulse excitation and echo reception, the motion of scattering particles is predominantly governed by hydrodynamic forces, with negligible influence from acoustic radiation effects; (2) scatterers are considered to exhibit perfect compliance with the local fluid velocity field.\\
In this work, PROTEUS was used to simulate particles trajectories within the vascular flow fields. The procedure followed was:

\begin{enumerate}
    \item {Import geometry and flow data:}  
    A new folder was created under \texttt{geometry\_data}, into which the CFD-generated flow field file (\texttt{.vtu}) was placed. The MATLAB script \texttt{vtu2matlab.m} was then modified to reference the subfolder and file name. Running this script converted the raw VTU into a \texttt{vtu.mat} file and automatically generated an STL representation and auxiliary data required by the simulator.
    
    \item {Preprocessing for inlet points:}  
    To define physiologically realistic injection sites for particles, the backpropagation procedure was executed. Specifically, I ran \texttt{backpropagation.m} in the \texttt{streamline-module} to propagate tracer particles backwards from the vasculature to the inlet plane using the flow velocity field. The resulting \texttt{backpropagation\_points.mat} file contained the streamline endpoints. A filtering step (\texttt{filter\_points.m}) was then applied to retain only those points lying on the inlet surface, resulting in \texttt{inlet\_points.mat}.
    
    \item {Probability density mapping:}  
    To translate the discrete inlet positions into a continuous injection distribution, I used a statistical density map of injection locations generated by Paraview, ensuring that particles seeding matched the flow characteristics.
    
    \item {Simulation and visualization:}  
    With inlet seeds defined, PROTEUS was used to initialize particle trajectories with randomized positions and sizes following the generated density distribution. Flow velocities imported from Musubi simulations \cite{2014End,2014Complex} were applied. I then visualized the resulting pathlines and spatial distributions of particless to study their transport behavior in the vascular network.
\end{enumerate}

\section{Generation of scatterers} \label{app:appD}
 When an ultrasound pulse excites the medium, each scatterer is treated as an independent monopole acoustic source, and its corresponding echo signal can be expressed as:

\begin{equation}
\begin{aligned}
P_{m}^{se}(\omega,t) 
&\approx \sum_{s=1}^{S} R_{s} \left[
    \sum_{n=1}^{vN}
        W_{n}
        \frac{e^{ikr_{ns}}}{r_{ns}}
        D(\theta_{ns},k)
        \delta(Y_{s}, r_{ns}, k)
        e^{i\omega\Delta\tau_n}
 \right] 
\\
&\qquad \times
    \frac{e^{ikr_{ms}}}{r_{ms}}
    D(\theta_{ms},k)
    \delta(Y_{s}, r_{ms}, k)
\end{aligned}
\label{eq:1}
\end{equation}

In equation \ref{eq:1}, $R_{s}$ denotes the reflection coefficient of the $s-th$ scatterer, representing local acoustic impedance differences. $W_{n}$ corresponds to the transmit aperture weighting function. The term $e^{ikr}$ describes spherical wave propagation attenuation ($k=2\pi f/c$ is the wavenumber, $r$ the propagation distance). $D(\theta,k)=\text{sinc}(kb\sin\theta)$ defines the element directivity function ($b$: half-width of the transducer element, $\theta$: azimuthal angle). $\delta(Y_{s},r,k)$ models elevation focusing effects using a Gaussian superposition method. $e^{i\omega\Delta\tau_{n}}$ implements delay control for transmit beamforming.
The pressure wave $P_{e}^{nv}(\omega,t)$ received by a transducer element is derived from the coherent summation of pressures across all its sub-elements. For the $nv-th$ element:
\begin{equation}
P_{e}^{nv}(\omega, t) = \sum_{\mu=0}^{v-1} P_{se}^{n+\mu}(\omega, t)
\label{eq:2}
\end{equation}

All theoretical pressures are derived for a single angular frequency $\omega=2\pi f$. Full-band waveforms are obtained through frequency-domain summation followed by inverse fast Fourier transform (IFFT). This model relies on the weak scattering assumption (single scattering events and no mutual interactions) and frequency-independent reflection coefficients. Computational efficiency and acoustic equivalence are balanced by optimizing scatterer density, with a typical value of 10 scatterers per wavelength squared. For example, under the typical imaging preset of an L11-4v probe (128 elements,Verasonics,Kirkland WA, USA) with a wavelength of approximately $200\,\mu\mathrm{m}$ at the center frequency, approximately 2.5 million scatterers can be generated within an imaging region of $4\text{ cm} \times 1.5\text{ cm} \times 5\text{ cm}$.

After generating scatterers, we determine whether the scatterers are located inside the vessel by projecting delta functions onto the band-limited space of the voxel grid\cite{Wise2019_41}. The three-dimensional delta function $\delta^{3}(r,\xi)$, centered at $r=\xi$, can be approximated as follows:
\begin{equation}
b(r, \xi) = \prod_{p} b\left(r^{(p)}, \xi^{(p)}\right)
\label{eq:3}
\end{equation}
with
\begin{equation}
b\big(r^{(p)},\xi^{(p)}\big)
= \frac{
    \sin\big(\pi(x^{(p)}-\xi^{(p)})\big)
  }{
    N^{(p)} \, \Theta\left(\dfrac{\pi(x^{(p)}-\xi^{(p)})}{N^{(p)} \Delta x^{(p)}}\right)
  }
\label{eq:4}
\end{equation}

Here, the superscript $(p)$ denotes the spatial dimension, $N^{(p)}$ is the number of grid points along the $p-th$ dimension, and $\Delta x^{(p)}$ is the grid spacing in that dimension. $x^{(p)}$ and $\xi^{(p)}$ are the vector components of $r$ and $\xi$, respectively. For the function $\Theta$ when $N^{(p)}$ is even, $\Theta=\tan()$; when $N^{(p)}$ is odd, $\Theta=\sin()$.
A set of scatterers $\xi_{i}$ can be represented on a discrete grid as:

\begin{equation}
s_{j}(t) = \sum_{i=1}^{M} C_{i}\, b(r_{i}, \xi_{i})\, s(\xi_{i}, t)
\label{eq:5}
\end{equation}

where $s_{j}(t)$ is the source signal at the $j-th$ grid point, and $C_{i}$ is the quadrature weight with $C_{i}=1/A$, where $A$ is the surface area per integration point in grid units, and $s(\xi_{i},t)$ represents the velocity vector at $\xi_{i}$. Once the projection position is obtained, the relationship between the index of the point's position and those of the boundary points can be used to determine whether the point is located inside the vessel.

\section{Optimized PFIELD Ultrasound Simulator} \label{app:appE}

\begin{algorithm}[H]
\caption{RF Signal Composition with Tissue Motion Consideration}
\label{alg:rf-composition}
\SetAlgoLined
\KwIn{
  Number of frames $N_{frames}$,\\
  Tissue motion status
}
\KwOut{Composite RF signal $\mathbf{RF_{total}}$}

\textbf{1: Frame Processing} \\
\For{$i = 1$ \KwTo $N_{frames}$}{
  1.1: \If{tissue is static}{
    \If{$i == 1$}{
      1.1.1: Compute surrounding tissue RF: $\mathbf{RF_{phantom}}$
    }
    1.1.2: Calculate flow RF: $\mathbf{RF_{flow}} \gets \text{flowModel}(i)$
  }
  \Else{
    1.2.1: Compute dynamic tissue RF: $\mathbf{RF_{phantom}} \gets \text{dynamicModel}(i)$ \\
    1.2.2: Calculate flow RF: $\mathbf{RF_{flow}} \gets \text{flowModel}(i)$
  }
  
  1.3: Compose total RF: $\mathbf{RF_{total}}[i] = \mathbf{RF_{phantom}} + \mathbf{RF_{flow}}$
}
\end{algorithm}

\begin{algorithm}[H]
\small
\caption{Distributed 3D Acoustic Pressure Field Calculation}
\label{alg:dist-acoustic-calc}
\SetAlgoLined
\KwIn{
  Scattering point sets $\mathbf{x}, \mathbf{y}, \mathbf{z}$,\\ 
  TX delays, transducer and medium parameters
}
\KwOut{RF spectrum $\mathbf{RFspectrum}$}

\textbf{1: Partition Phase} \\
1.1: Initialize GPU and estimate required memory per block \\
1.2: Determine number of blocks $NW$ \\
1.3: Partition input data into $NW$ chunks: \\
\hspace{1.5em}Divide $\mathbf{x}, \mathbf{y}, \mathbf{z}, \mathbf{RC}$ into sublists \\
\hspace{1.5em}$\mathbf{x}_i, \mathbf{y}_i, \mathbf{z}_i, \mathbf{RC}_i$ $(i=1,...,NW)$ \\

\textbf{2: Distributed Computation Phase} \\
\For{$i = 1$ \KwTo $NW$}{
  2.1: \textbf{Acoustic field computation on block }$i$ \\
  2.2: Subdivide each transducer element into sub-elements \\
  2.3: Precompute geometric distances and angles between \\
  \hspace{1.5em}field points $(\mathbf{x}_i,\mathbf{y}_i,\mathbf{z}_i)$ and sub-elements \\
  2.4: Determine frequency samples and prepare spectral weights \\
  2.5: Pre-allocate required GPU arrays \\
  
  2.6: \For{each frequency sample $f_k$}{
    2.6.1: Compute wavenumber $k_w = 2\pi f_k / c$ and attenuation \\
    2.6.2: For all field points and sub-elements: \\
    \hspace{2.5em}Compute $r$, $1/r$, and necessary corrections \\
    2.6.3: Update GPU arrays with propagation/attenuation \\
    2.6.4: For each array element: \\
    \hspace{2.5em}Sum over sub-elements with delay and apodization \\
    2.6.5: Sum element contributions at each field point: \\
    \hspace{2.5em}$RP_k \gets \textrm{sum of contributions}$ \\
    2.6.6: Multiply $RP_k$ by spectral weights and accumulate: \\
    \hspace{2.5em}$RP_i \gets RP_i + RP_k \times \textrm{weights}$
  }
  
  2.7: Integrate $RP_i$ over frequencies for RMS field \\
  2.8: Store result $RFspD_i$ and indices $\mathrm{idx}_i$ \\
}

\textbf{3: Combination Phase} \\
3.1: Initialize $\mathbf{RFsp} \gets \mathbf{0}$ \\
3.2: \For{$i = 1$ \KwTo $NW$}{
  $\mathbf{RFsp} \gets \mathbf{RFsp} + RFspD_i$
}
3.3: Assign $\mathbf{RFspectrum} \gets \mathbf{RFsp}$ using indices $\mathrm{idx}_i$ \\
\end{algorithm}

\section{uPDI-optimized reconstructor} \label{app:appF}

\begin{algorithm}[H]
\small
\caption{Memory-Efficient Delay-and-Sum Beamforming via Chunk Processing}
\label{alg:alg3}
\SetAlgoLined
\KwIn{
  RF signals per frame: $\mathbf{RF}$, 
  Spatial coordinates: $\mathbf{x}, \mathbf{y}, \mathbf{z}$,\\
  Transmit events: $N_a$, 
  Total frames: $N_{frames}$, 
  Available memory: $M_{PAB}$
}
\KwOut{
  Reconstructed 3D complex volumes: $\{\mathbf{Vol}_i\}_{i=1}^{N_{frames}}$
}

\textbf{1: Initialization Phase} \\
\Indp
1.1: Calculate total spatial points $N_{points} \gets \text{len}(\mathbf{x})$ \\
1.2: Estimate memory requirement: \\
\hspace{1.5em}$bytes \gets 16 \times N_{points} \times N_a$ \\
1.3: Determine chunks: $N_{chunks} \gets \lceil bytes / M_{PAB} \rceil$ \\
1.4: Partition spatial grid into $N_{chunks}$ contiguous blocks \\
\Indm

\textbf{2: Chunk-wise Beamforming} \\
\For{$chunk_i = 1$ \KwTo $N_{chunks}$}{
  2.1: Extract subgrid coordinates: \\
  \hspace{1.5em}$\mathbf{x}_i \gets \mathbf{x}[idx_i\!:\!idx_{i+1}\!-\!1]$, 
  $\mathbf{y}_i \gets \mathbf{y}[idx_i\!:\!idx_{i+1}\!-\!1]$, 
  $\mathbf{z}_i \gets \mathbf{z}[idx_i\!:\!idx_{i+1}\!-\!1]$ \\
  
  2.2: \For{$tx_k = 1$ \KwTo $N_a$}{
    2.2.1: Compute delay matrix: \\
    \hspace{1.5em}$\mathbf{M}_{tx_k} \gets \text{dasmtx3}(\mathbf{x}_i, \mathbf{y}_i, \mathbf{z}_i, \text{tx\_params})$
  }
  
  2.3: \For{$frame_j = 1$ \KwTo $N_{frames}$}{
    2.3.1: Load raw RF data: $\mathbf{SIG} \gets \mathbf{RF}_{frame_j}$ \\
    
    2.3.2: \For{$tx_j = 1$ \KwTo $N_a$}{
      2.3.2.1: Convert to IQ: $\mathbf{b} \gets \text{rf2iq}(\mathbf{SIG}_{tx_j}, f_s, f_c)$ \\
      2.3.2.2: Zero-pad: $\widetilde{\mathbf{SIG}} \gets [\mathbf{b}; \mathbf{0}_{pad}]$ \\
      2.3.2.3: Vectorize: $\mathbf{sigVec} \gets \text{vec}(\widetilde{\mathbf{SIG}})$ \\
      2.3.2.4: Apply delays: \\
      \hspace{2.5em}$\mathbf{IQ}_{chunk}(:,frame_j,tx_j) \gets \mathbf{M}_{tx_j} \times \mathbf{sigVec}$
    }
  }
  
  2.4: Average over transmits: $\mathbf{IQ}_{sum} \gets \text{mean}(\mathbf{IQ}_{chunk}, 3)$ \\
  2.5: Save $\mathbf{IQ}_{sum}$ to $\text{IQ\_CHUNK}_{chunk_i}.mat$ \\
}

\textbf{3: Frame-wise Volume Reconstruction} \\
\For{$frame_i = 1$ \KwTo $N_{frames}$}{
  3.1: Initialize empty buffer: $\mathbf{dataList} \gets [\ ]$ \\
  
  3.2: \For{$chunk_k = 1$ \KwTo $N_{chunks}$}{
    3.2.1: Load $\mathbf{IQ}_{sum} \gets \text{IQ\_CHUNK}_{chunk_k}.mat$ \\
    3.2.2: Extract frame data: $\mathbf{dataChunk} \gets \mathbf{IQ}_{sum}(:, frame_i)$ \\
    3.2.3: Concatenate: $\mathbf{dataList} \gets [\mathbf{dataList}; \mathbf{dataChunk}]$
  }
  
  3.3: Reshape to 3D volume: $\mathbf{Vol}_i \gets \text{reshape}(\mathbf{dataList})$ \\
  3.4: Save $\mathbf{Vol}_i$ to $\text{Frame}_{frame_i}.mat$ \\
}
\end{algorithm}

%% file: main.bbl
\begin{thebibliography}{10}
\expandafter\ifx\csname url\endcsname\relax
  \def\url#1{\texttt{#1}}\fi
\expandafter\ifx\csname urlprefix\endcsname\relax\def\urlprefix{URL }\fi
\expandafter\ifx\csname href\endcsname\relax
  \def\href#1#2{#2} \def\path#1{#1}\fi

\bibitem{2014Ultrafast}
M.~Tanter, M.~Fink, Ultrafast imaging in biomedical ultrasound, Ultrasonics Ferroelectrics \& Frequency Control IEEE Transactions on 61~(1) (2014) 102--119.

\bibitem{2017Noninvasive}
D.~Maresca, M.~Correia, O.~Villemain, A.~Bizé, L.~Sambin, M.~Tanter, B.~Ghaleh, M.~Pernot, Noninvasive imaging of the coronary vasculature using ultrafast ultrasound, Jacc Cardiovascular Imaging (2017).

\bibitem{2013Functional}
E.~Mace, G.~Montaldo, B.~F. Osmanski, I.~Cohen, M.~Tanter, Functional ultrasound imaging of the brain: theory and basic principles, IEEE Transactions on Ultrasonics, Ferroelectrics, and Frequency Control 60~(3) (2013) 492--506.

\bibitem{2015Spatiotemporal}
C.~Demene, T.~Deffieux, M.~Pernot, B.~F. Osmanski, V.~Biran, J.~L. Gennisson, L.~A. Sieu, A.~Bergel, S.~Franqui, J.~M. Correas, Spatiotemporal clutter filtering of ultrafast ultrasound data highly increases doppler and fultrasound sensitivity, IEEE Transactions on Medical Imaging 34~(11) (2015) 2271--2285.

\bibitem{2011Ultrafast}
J.~Bercoff, G.~Montaldo, T.~Loupas, D.~Savery, F.~Meziere, M.~Fink, M.~Tanter, Ultrafast compound doppler imaging: providing full blood flow characterization, Ultrasonics Ferroelectrics \& Frequency Control IEEE Transactions on 58~(1) (2011) 134--147.

\bibitem{2025In}
H.~Yu, Zhang, F.~J., Feng, J.~Yang, Y.~Xia, Y.~Zhao, J.~Zhang, In vivo dynamic coronary arteries blood flow imaging based on multi-cycle phase clustering ultrafast ultrasound, Advanced science 12 (2025).

\bibitem{0Improved}
U.~W. Lok, J.~D. Trzasko, C.~Huang, J.~Zhang, R.~M. Deruiter, S.~Chen, Improved curvelet transform-based sparsity promoting algorithm for fast ultrasound localization microscopy, in: 2024 IEEE Ultrasonics, Ferroelectrics, and Frequency Control Joint Symposium (UFFC-JS).

\bibitem{Belgharbi2023_19}
H.~Belgharbi, J.~Porée, R.~Damseh, V.~Perrot, L.~Milecki, P.~Delafontaine-Martel, F.~Lesage, J.~Provost, An anatomically realistic simulation framework for 3d ultrasound localization microscopy, IEEE Open Journal of Ultrasonics, Ferroelectrics, and Frequency Control 3 (2023) 1--13.

\bibitem{Garcia2022_20}
D.~Garcia, Simus: An open-source simulator for medical ultrasound imaging. part i: Theory \& examples, Computer Methods and Programs in Biomedicine 218 (2022) 106726.

\bibitem{Ekroll2023_21}
I.~Ekroll, A.~Saris, J.~Avdal, Flust: A fast, open source framework for ultrasound blood flow simulations, Computer Methods and Programs in Biomedicine 238~(000) (2023) 13.

\bibitem{Heiles2025_22}
B.~Heiles, A.~Chavignon, V.~Hingot, P.~Lopez, E.~Teston, O.~Couture, Performance benchmarking of microbubble-localization algorithms for ultrasound localization microscopy, Nature biomedical engineering 6 (2022) 605.

\bibitem{Lerendegui2022_23}
M.~Lerendegui, K.~Riemer, B.~Wang, C.~Dunsby, M.-X. Tang, Bubble flow field: A simulation framework for evaluating ultrasound localization microscopy algorithms, arXiv preprint arXiv:2211.00754 (2022).

\bibitem{Jensen1992_24}
J.~Jensen, N.~Svendsen, Calculation of pressure fields from arbitrarily shaped, apodized, and excited ultrasound transducers, IEEE Transactions on Ultrasonics, Ferroelectrics, and Frequency Control 39~(2) (1992) 262--267.

\bibitem{Jensen1999_25}
J.~Jensen, Field: A program for simulating ultrasound systems (1999).

\bibitem{BlankenXX_26}
N.~Blanken, B.~Heiles, A.~Kuliesh, M.~Versuis, K.~Jain, D.~Maresca, G.~Lajoinie, Proteus: A physically realistic contrast-enhanced ultrasound simulator---part i: Numerical methods, IEEE Transactions on Ultrasonics, Ferroelectrics, and Frequency Control PP (2024).

\bibitem{Treeby2010_27}
B.~Treeby, B.~Cox, k-wave: Matlab toolbox for the simulation and reconstruction of photoacoustic wave fields, Journal of biomedical optics 15~(2) (2010) 021314.

\bibitem{2022Super}
N.~Blanken, J.~M. Wolterink, H.~Delingette, C.~Brune, M.~Versluis, G.~Lajoinie, Super-resolved microbubble localization in single-channel ultrasound rf signals using deep learning, IEEE Transactions on Medical Imaging (2022).

\bibitem{2021Super}
R.~J.~G. Van~Sloun, O.~Solomon, M.~Bruce, Z.~Z. Khaing, H.~Wijkstra, Y.~C. Eldar, M.~Mischi, Super-resolution ultrasound localization microscopy through deep learning, IEEE Transactions on Medical Imaging~(3) (2021) 40.

\bibitem{No0Functional}
N.~Renaudin, C.~Demené, A.~Dizeux, N.~Ialy-Radio, S.~Pezet, M.~Tanter, Functional ultrasound localization microscopy reveals brain-wide neurovascular activity on a microscopic scale, Nature Methods 19~(8) (2022) 1004--1012.

\bibitem{2024Pruning}
B.~Rauby, P.~Xing, J.~Porée, M.~Gasse, J.~Provost, Pruning sparse tensor neural networks enables deep learning for 3d ultrasound localization microscopy, IEEE Trans Image Process 34 (2025) 2367--2378.

\bibitem{Shin2024_18}
M.~A.~A. YiRang~Shin, P.~Song, Context-aware deep learning enables high-efficacy localization of high concentration microbubbles for super-resolution ultrasound localization microscopy, Nat Commun 15~(1) (2024) 2932.

\bibitem{GalarretaValverde2013_28}
M.~A. Galarreta-Valverde, M.~M.~G. Macedo, C.~Mekkaoui, M.~P. Jackowski, S.~Ourselin, D.~R. Haynor, Three-dimensional synthetic blood vessel generation using stochastic l-systems, in: SPIE Medical Imaging, Vol. 8669, SPIE, 2013.

\bibitem{Adiv1985_30}
Adiv, Gilad, Determining three-dimensional motion and structure from optical flow generated by several moving objects, Pattern Analysis \& Machine Intelligence IEEE Transactions on (1985).

\bibitem{Updegrove2017_31}
A.~Updegrove, N.~M. Wilson, J.~Merkow, H.~Lan, A.~L. Marsden, S.~C. Shadden, Simvascular: An open source pipeline for cardiovascular simulation, Annals of Biomedical Engineering 45~(3) (2017) 1--17.

\bibitem{Lan2017_32}
H.~Lan, A.~Updegrove, N.~Wilson, G.~D. Maher, S.~C. Shadden, A.~Marsden, A re-engineered software interface and workflow for the open source simvascular cardiovascular modeling package, J Biomech Eng (2017).

\bibitem{Taylor1998_33}
C.~Taylor, T.~Hughes, C.~Zarins, Finite element modeling of blood flow in arteries, Computer Methods in Applied Mechanics \& Engineering 158~(1-2) (1998) 155--196.

\bibitem{Marsden2015_34}
A.~L. Marsden, M.~Esmaily-Moghadam, Multiscale modeling of cardiovascular flows for clinical decision support, Applied Mechanics Reviews: An Assessment of the World Literature in Engineering Sciences 67~(1/6) (2015) 030804--1.

\bibitem{Sahni2008_35}
O.~Sahni, K.~E. Jansen, M.~S. Shephard, C.~A. Taylor, M.~W. Beall, Adaptive boundary layer meshing for viscous flow simulations, Engineering with Computers 24~(3) (2008) 267--285.

\bibitem{Christian2001_36}
C.~H. Whiting, K.~E. Jansen, A stabilized finite element method for the incompressible navier--stokes equations using a hierarchical basis, International Journal for Numerical Methods in Fluids 35~(1) (2001) 93--116.

\bibitem{2005ParaView_39}
J.~Ahrens, B.~Geveci, C.~Law, Paraview: An end-user tool for large data visualization, Visualization Handbook (2005) 717--731.

\bibitem{Ayachit2015_40}
U.~Ayachit, The ParaView Guide: A Parallel Visualization Application, Kitware, Inc., 2015.

\bibitem{1997The}
L.~F. Shampine, M.~W. Reichelt, The matlab ode suite, Siam Journal on Scientific Computing 18~(1) (1997) 1--22.

\bibitem{2018Meshfree}
S.~Shahriari, D.~Garcia, Meshfree simulations of ultrasound vector flow imaging using smoothed particle hydrodynamics, Physics in medicine and biology 63~(20) (2018) 205011.

\bibitem{2009Assessment}
A.~Swillens, T.~D. Schryver, L.~L?Vstakken, H.~Torp, P.~Segers, Assessment of numerical simulation strategies for ultrasonic color blood flow imaging, based on a computer and experimental model of the carotid artery, Annals of Biomedical Engineering 37~(11) (2009) 2188--2199.

\bibitem{2016High}
J.~Poree, D.~Posada, A.~Hodzic, F.~Tournoux, G.~Cloutier, D.~Garcia, High-frame-rate echocardiography using coherent compounding with doppler-based motion-compensation, IEEE Trans Med Imaging 35~(7) (2016) 1647--1657.

\bibitem{2018Color}
C.~Madiena, J.~Faurie, J.~Poree, D.~Garcia, Color and vector flow imaging in parallel ultrasound with sub-nyquist sampling., IEEE Transactions on Ultrasonics Ferroelectrics \& Frequency Control (2018) 795--802.

\bibitem{GARCIA2024108169}
D.~Garcia, F.~Varray, \href{https://www.sciencedirect.com/science/article/pii/S0169260724001652}{Simus3: An open-source simulator for 3-d ultrasound imaging}, Computer Methods and Programs in Biomedicine 250 (2024) 108169.
\newblock \href {https://doi.org/https://doi.org/10.1016/j.cmpb.2024.108169} {\path{doi:https://doi.org/10.1016/j.cmpb.2024.108169}}.
\newline\urlprefix\url{https://www.sciencedirect.com/science/article/pii/S0169260724001652}

\bibitem{6841027}
G.~Y. Hou, J.~Provost, J.~Grondin, S.~Wang, F.~Marquet, E.~Bunting, E.~E. Konofagou, Sparse matrix beamforming and image reconstruction for 2-d hifu monitoring using harmonic motion imaging for focused ultrasound (hmifu) with in vitro validation, IEEE Transactions on Medical Imaging 33~(11) (2014) 2107--2117.
\newblock \href {https://doi.org/10.1109/TMI.2014.2332184} {\path{doi:10.1109/TMI.2014.2332184}}.

\bibitem{2020So}
V.~Perrot, M.~Polichetti, F.~Varray, D.~Garcia, So you think you can das? a viewpoint on delay-and-sum beamforming, Ultrasonics (2020).

\bibitem{9593605}
D.~Garcia, Make the most of must, an open-source matlab ultrasound toolbox, in: 2021 IEEE International Ultrasonics Symposium (IUS), 2021, pp. 1--4.
\newblock \href {https://doi.org/10.1109/IUS52206.2021.9593605} {\path{doi:10.1109/IUS52206.2021.9593605}}.

\bibitem{Troianowski2011_38}
Taylor, A.~C., Three-dimensional simulations in glenn patients: Clinically based boundary conditions, hemodynamic results and sensitivity to input data, Journal of Biomechanical Engineering 133~(11) (2011).

\bibitem{2023Interactive}
J.~Xu, G.~Thevenon, T.~Chabat, M.~Mccormick, F.~Li, T.~Birdsong, K.~Martin, Y.~Lee, S.~Aylward, Interactive, in-browser cinematic volume rendering of medical images, Computer Methods in Biomechanics and Biomedical Engineering: Imaging And Visualization 11~(4) (2023) 1019--1026.

\bibitem{Si2015_37}
Si, Hang, Tetgen, a delaunay-based quality tetrahedral mesh generator, Acm Transactions on Mathematical Software 41~(2) (2015) 1--36.

\bibitem{2014End}
H.~Klimach, K.~Jain, S.~Roller, End-to-end parallel simulations with apes, in: Parallel Computing: Accelerating Computational Science and Engineering (CSE), 2014.

\bibitem{2014Complex}
M.~Hasert, K.~Masilamani, S.~Zimny, H.~Klimach, J.~Qi, J.~Bernsdorf, S.~Roller, Complex fluid simulations with the parallel tree-based lattice boltzmann solver musubi, Journal of Computational Science 5~(5) (2014) 784--794.

\bibitem{Wise2019_41}
E.~S. Wise, B.~T. Cox, J.~Jaros, B.~E. Treeby, Representing arbitrary acoustic source and sensor distributions in fourier collocation methods, The Journal of the Acoustical Society of America 146~(1) (2019) 278--288.

\end{thebibliography}
